# Vertical coupling of laser glass microspheres to buried silicon nitride ellipses and waveguides

D. Navarro-Urrios[1,a)], J. M. Ramírez[2], N.E. Capuj[3], Y. Berencén[2,†], B. Garrido[2] and A. Tredicucci[4]

[1] NEST, Istituto Nanoscienze – CNR and Scuola Normale Superiore, Piazza San Silvestro 12, Pisa, I-56127, Italy

[2] Departament d'Electrònica, Universitat de Barcelona, Barcelona 08028, Spain

[3] Depto. Física, Universidad de la Laguna, 38206, Spain

[4] NEST, Istituto Nanoscienze and Dipartimento di Fisica, Università di Pisa, Largo Pontecorvo 3, I-56127 Pisa, Italy

*We demonstrate the integration of $Nd^{3+}$ doped Barium-Titanium-Silicate microsphere lasers with a Silicon Nitride photonic platform. Devices with two different geometrical configurations for extracting the laser light to buried waveguides have been fabricated and characterized. The first configuration relies on a standard coupling scheme, where the microspheres are placed over strip waveguides. The second is based on a buried elliptical geometry whose working principle is that of an elliptical mirror. In the latter case, the input of a strip waveguide is placed on one focus of the ellipse, while a lasing microsphere is placed on top of the other focus. The fabricated elliptical geometry (ellipticity=0.9) presents a light collecting capacity that is 50% greater than that of the standard waveguide coupling configuration and could be further improved by increasing the ellipticity. Moreover, since the dimensions of the spheres are much smaller than those of the ellipses, surface planarization is not required. On the contrary, we show that the absence of a planarization step strongly damages the microsphere lasing performance in the standard configuration.*

---

[a] Author to whom correspondence should be addressed.  Electronic mail: daniel.navarrourrios@nano.cnr.it.

[†] Present address: Institute of Ion Beam Physics and Materials Research, Helmholtz-Zentrum Dresden-Rossendorf, P.O. Box 510119, 01314 Dresden, Germany

## I. INTRODUCTION

The most efficient strategy for extracting light on-chip from whispering gallery mode (WGM) lasing resonators such as disks, rings or toroids is to exploit the evanescent coupling to bus waveguides [1-4]. However, in spherical or ellipsoid geometries this is not necessarily true, since the energy exchange is restricted to the limited set of WGMs with propagating vectors below the light line of the waveguide [5]. This is not an issue if the pumping light is coupled-in through the same bus waveguide [6-7], but in a far-field pumping configuration [8-9] a sizeable fraction of the emitted light is unavoidably lost. A direct consequence when aiming for sensing application [10] is that, although WGM lasing microspheres provide a high interaction surface, most of it is unexploited.

Evanescent coupling to tapered optical fibers is the standard way of exciting and collecting laser light from WGM resonators, but those fibers are extremely fragile and sensible to the environment conditions [11]. In the case of microspheres, vertical coupling to buried waveguides is the straightforward configuration to achieve some degree of integration and out-of-the-laboratory applicability. Those waveguides can be part of more complex photonic circuits with additional functionalities. The use of buried planar lightwave circuits provides further benefits, since the vertical coupling approach allows controlling the gap width with extreme precision and the selective exposure of the resonator to the environment for sensing applications. The first demonstration of an integrated system of this kind was reported by Murugan et al. [12], where $Nd^{3+}$-doped borosilicate microsphere lasers were successfully coupled to channel waveguides fabricated by ion exchange.

In the present work, we demonstrate a novel coupling strategy based on the vertical evanescent coupling to a buried $Si_3N_4$ ellipse, which enables a 50% enhancement of the light collecting capacity when compared to vertically-coupled buried waveguides made of the same material. We have exploited one of the most interesting properties of the ellipse in optics, being that when a ray of light originating from one focus reflects off its inner surface, it always passes through the other focus. We have placed a lasing sphere over one of the foci and the input of a strip waveguide in the other, so that light emitted from the microsphere is focused on the entrance of the waveguide. This configuration integrates a range of azimuthal angles that satisfy the condition of total internal reflection and lie below the light line of the waveguide.

## II. FABRICATION

One of the advantages of using glass microsphere resonators is the material flexibility, both in the glass composition and in the dopant specie. In this work, the microspheres were fabricated from Barium-Titanium-Silicate (BTS) glass doped with $Nd^{3+}$ ions with the composition of 40%BaO–20%$TiO_2$–40%$SiO_2$ and doped with 1.5% $Nd_2O_3$ (in the molar ratio). The glass is reduced to dust by means of a mortar and is heated up to its fusion temperature, which is around 900°C. Most of the splinters melt and, when the temperature decreases, solidify in a spherical shape of tens of micrometers [13] (see Figure 1 for a SEM micrograph of one of the fabricated microspheres). Details of the light emission properties of these microspheres can be found elsewhere [8-9].

The photonic platform was fabricated using standard CMOS processes (Figure 1). As a first step, a 2 μm thick $SiO_2$ layer was thermally grown on top of a crystalline silicon wafer. A 0.5 μm thick layer of stoichiometric $Si_3N_4$ (refractive index $n_1$=2) was then deposited using the low-pressure chemical vapor deposition technique. The thickness of this layer was chosen to ensure monomodal behavior in the vertical direction at 1μm. Subsequently, different photonic geometries were defined by standard photolithographic techniques. The strip waveguides are 5 μm wide and the ellipses have an ellipticity e=c/a=0.9 (where c=650 μm is the distance from the center to a focus, a=725 μm is the major semi-axis and b is the minor semi-axis), which were defined and then covered by another $SiO_2$ cladding layer (refractive index $n_2$=1.45) of 0.65 μm. No mechanical polishing was applied to the surface of the resulting layer. Finally, 0.5 μm deep circular holes were opened in the top layer. Different hole diameters (w) were fabricated to perfectly fit microspheres in correspondence to their radii. When inserted in the hole, the microspheres are aligned in one case with respect to a buried waveguide and in the other to the focal point of an ellipse (see Figure 1 for an optical image of one of the ellipses). The vertical gap between the top of the $Si_3N_4$ layer and the bottom part of the microsphere is about 0.15 μm. Measurements done with a profilometer revealed a protuberance on the floor of the circular opening, which is associated to the presence of the buried waveguide. This effect is obviously not observed on holes placed on top of an ellipse, where the profile of the hole bottom is completely flat.

The optical losses of the $Si_3N_4$ waveguides were assessed to be less than 0.8 dB/cm at 780 nm (about the maximum sensitivity of our setup) [14], showing a monotonic decreasing behavior with wavelength.

The microspheres were deposited over the photonic platform and driven into the circular openings by gently pushing them with micrometric needles mounted on a micro-precision stage. Once inserted in the holes, the microspheres are stable enough to allow the wafer physical transport even without covering the surface. The samples were cut in a way that the distance from the sample edge to the microsphere was about 3mm.

### III. EXPERIMENTAL SETUP

For the optical measurements (see Figure 2), a continuous wave laser diode at 808 nm is focused on a spot of $1.5 \times 10^{-5}$ $cm^2$ using an infrared objective (O1). A single microsphere can be aligned with the pump beam using a xyz micro-precision stage, the spot size being large enough to guarantee homogeneous spatial pumping over the microsphere volume. Indeed, the spatial profile of the photoluminescence emission looks spatially homogeneous (see optical image of Figure 2). The emitted signal is collected by another objective (O2) placed in front of the microsphere. Since the working distance of O2 is about 4 mm, i.e., bigger that the distance from the sample edge to the microsphere, it is possible to have the collection focus either at the microsphere position (Focal Plane Sphere, FPS from now on) or at the waveguide output (Focal Plane Waveguide, FPW from now on). Both focal planes are illustrated in Figure 1. The diameter of the microspheres is far greater than the spatial resolution of the collection system (about 5 μm) [15]. Thus, when O2 is at FPS, the collected emission comes out from a localized volume, which was specifically chosen to be close to the north-pole of the microsphere. A long-pass filter with a cut-off wavelength of 950 nm is used to filter out the laser beam excitation. Collected light is then focused at the entrance slit of a 750 mm monochromator. For signal detection, a CCD is placed at one of the monochromator output ports and directly connected to a computer. The dimensions of the microspheres are quantified by forming the device image in the CCD and using a structure of known size as calibration. In this work, we have used nearly equivalent microspheres with radii R≈20µm.

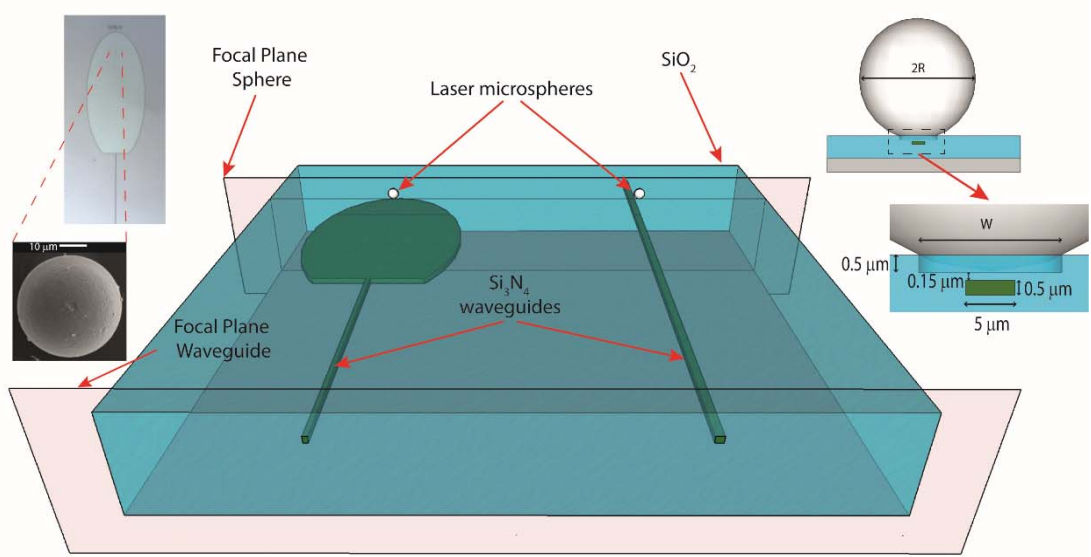

**Figure 1.** Scheme of the fabricated structures and the two possible focal planes (Focal Plane Waveguide and Focal Plane Sphere) related to the microscope O2 position. On the left, an optical image of an ellipse (notice the hole in the top focus) and a SEM micrograph of one of the fabricated microspheres. On the right, a geometrical cross section of the nominal coupled geometry (showing a buried waveguide in this case) with the characteristic dimensions.

The significant transitions among the energy levels of $Nd^{3+}$ ions are reported in Figure 2. The incoming laser pumps the ions from the ground state ($^4I_{9/2}$) to the $^4F_{5/2}$ and $^2H_{9/2}$ levels, from where a rapid non-radiative thermalization to the $^4F_{3/2}$ level takes place. Population inversion is achieved between the $^4F_{3/2}$ and the $^4I_{11/2}$ states, whose radiative transition is well known to be very efficient [16]. Finally, the system thermalizes to the ground state. Hence, this is effectively a four-level system where $N_4$, $N_3$, $N_2$ and $N_1$ correspond to the number of ions populating the $^4F_{5/2}$, $^4F_{3/2}$, $^4I_{11/2}$ and $^4I_{9/2}$ levels respectively. Laser action at around 1064 nm occurs in a given eigenmode of the WGM cavity if $N_3 > N_2$ and the passive losses of the specific trajectory are compensated. In this work, we mainly deal with modes that lie within vertical planes containing the north and south poles of the spheres. Collected light comes out of the microspheres from one of the poles with a propagating vector angle equal to the azimuthal angle of the plane containing the particular WGM. The real and imaginary part of the effective refractive index of a given WGM depend on the geometry of the cavity and the optical losses. The latter are mainly associated to scattering [15] when direct absorption from the ion species is negligible, which is the case of the four-level system under study. In our microspheres, losses are dominated by the local characteristics of the microsphere surface and volume. Provided that the pumping source is spatially homogeneous, the previous combined characteristics allow lasing action only for specific ranges of azimuthal angles where losses are low enough. In addition, lasing modes associated to different azimuthal angles appear at different wavelengths due to the slight asphericity of the resonators and to small local variations of the gain spectrum.

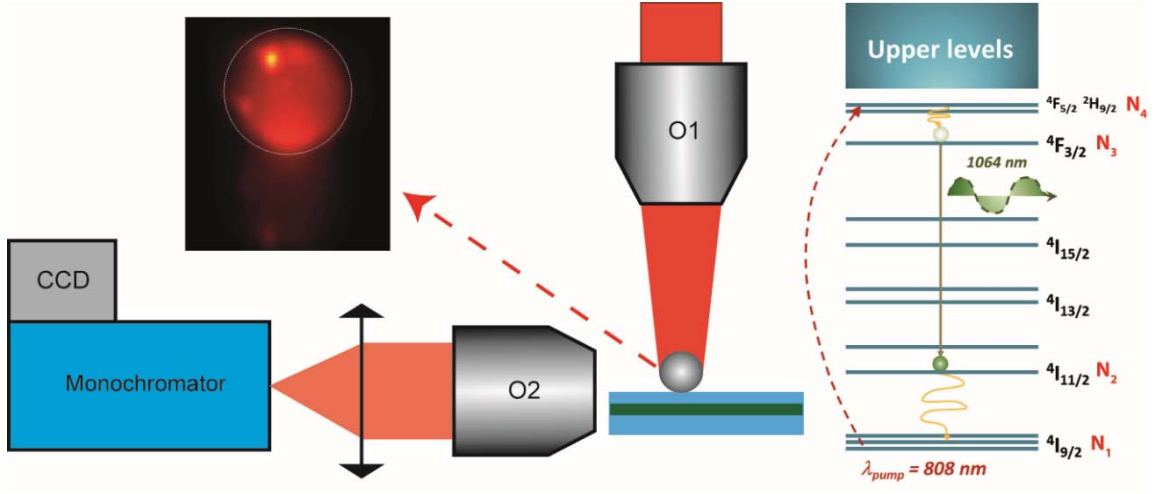

**Figure 2.** Scheme of the optical setup. The image on the left shows the spatial profile of the emitted PL below threshold, which has been taken directly from the microscope using a filter that eliminates the scattering from the pumping laser. The sphere contour is highlighted for clarity. The bright spot is associated to a localized scattering region within the sphere. Note that some signal comes from below the sphere, which is due to reflection on the substrate. On the right, a schematic diagram of $Nd^{3+}$ energy levels under excitation at 808 nm.

**IV. LIGHT COLLECTING CAPACITY: ELLIPSE VS WAVEGUIDE**

In order to quantify the light collecting capacity of the ellipse configuration, we first define $\Theta_1$ as the polar angle with respect to the focus in which the microsphere is placed. $\Theta_1$ can also be interpreted as the azimuth angle of a given WGM of the microsphere contained in a vertical plane perpendicular to the substrate. The reference direction is taken towards the center of the ellipse origin. Thus, starting from the ellipse formula in polar coordinates with respect to a focus ($r_i(\Theta_i) = \dfrac{a(1-e^2)}{1-e\cos(\Theta_i)}$, $r_i$ being the distance to the ellipse surface as measured from the focus i=1,2) and using the relation $r_1 + r_2 = 2a$, it is straightforward to write an expression linking $\Theta_1$ and $\Theta_2$ (see top scheme of Figure 3):

$$\cos(\Theta_1) = \dfrac{2e - \cos(\Theta_2)(e^2+1)}{1 - 2e\cos(\Theta_2) + e^2} \qquad (1).$$

The minimum angle that is accepted by the waveguide ($\Theta_{1,wg}$) is calculated by imposing $\Theta_{2,wg} = \alpha_{wg} = \cos^{-1}\left(n_2/n_1\right)$ in Eq. 1:

$$\Theta_{1,wg} = \cos^{-1}\left(\dfrac{2e - (e^2+1)\dfrac{n_2}{n_1}}{1 - 2e\dfrac{n_2}{n_1} + e^2}\right) \qquad (2),$$

On the other hand, the ellipse will only behave as a perfect mirror below the minimum angle ($\Theta_{1,ir}$) ensuring total internal reflection within the ellipse:

$$\Theta_{1,ir} + \cos^{-1}\left(\frac{2e - \cos(\Theta_{1,ir})(e^2 + 1)}{1 - 2e\cos(\Theta_{1,ir}) + e^2}\right) + 2\cos^{-1}\left(n_2/n_1\right) = 0 \qquad (3).$$

Therefore, the optimum working range of the ellipse is $\Theta_{1,wg} < \Theta_1 < \Theta_{1,ir}$. Within this range, light emitted by the microsphere is totally reflected in the inner elliptical surface and focused at the entrance of the waveguide, where it propagates below the light line. In Figure 3, we have represented $\Theta_{1,wg}$ and $\Theta_{1,ir}$ as extracted from Eqs. 2 (green curve) and 3 (red curve) as a function of the ellipticity (we have used $Si_3N_4$ and $SiO_2$ as the high and low index materials respectively). The difference between those two angles ($\Theta_{1,ir}$-$\Theta_{1,wg}$) is represented as the black curve. We have also plotted the minimum propagation angle ensuring total internal reflection in the waveguide ($\alpha_{wg} = 43°$, dashed orange curve). The ellipse configuration starts collecting light above $e>0.72$ and beats the standard waveguide configuration for $e>0.82$, which occurs for $\Theta_{1,ir}$-$\Theta_{1,wg}$>$\alpha_{wg}$. For the fabricated devices (vertical dashed line), the light collecting capacity of the ellipse configuration 2*($\Theta_{1,ir}$-$\Theta_{1,wg}$) is 130°, which is about 50% greater than that of the waveguide (2*$\alpha_{wg}$=86°) and could reach 100% for even higher ellipticities. It is worth noting that the light collecting capacity increases with the refractive index contrast, and therefore a suitable selection of materials could provide better results. In our case, we have used $Si_3N_4$ due to the extremely low propagation losses in the spectral range covered by the $^4F_{3/2}$ -> $^4I_{11/2}$ transition of the $Nd^{3+}$ ions.

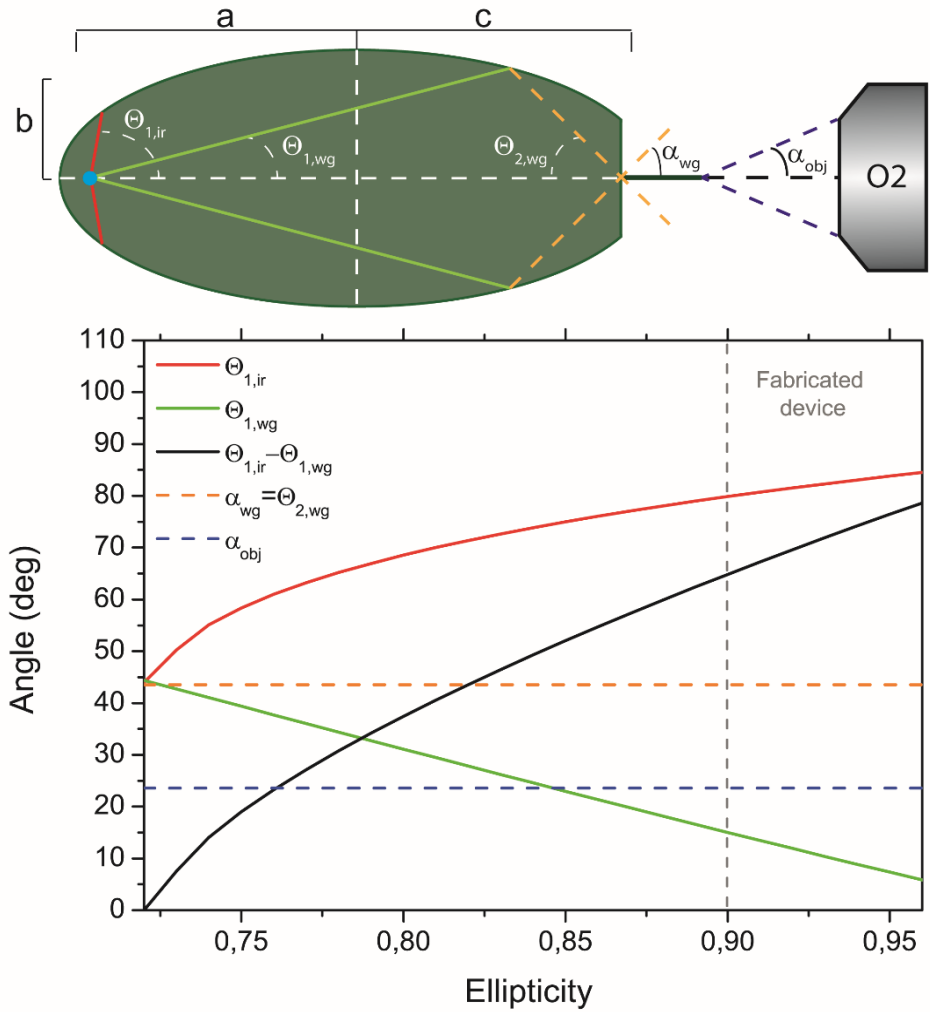

**Figure 3.** Calculation of the maximum internal reflection angle within the ellipse (red curve) and the minimum waveguide acceptance angle (green curve) referred to the focus where the microsphere is placed. The black curve represents $\Theta_{1,ir}-\Theta_{1,wg}$, which is half of the light collecting capacity of the ellipse. Horizontal dashed curves represent the acceptance angle in the buried waveguide configuration (orange) and the collection angle of microscope O2 (blue). The curves are represented as a function of the ellipticity. The top scheme depicts the previous angles referred to the ellipse configuration. Notice that the microscope O2 is not on scale.

**V. OPTICAL MEASUREMENTS**

As it can be observed in Figure 4 (panels a) and b)), both coupling geometries allow for laser light extraction through the waveguide. Two different lasing microspheres were studied in each configuration, all of them pumped at the same photon flux. In the case of direct coupling to the waveguide, the obtained spectra for the two possible focal positions of microscope O2 are very similar, since the range of collection angles starts from zero. On the other hand, in the ellipse configuration, lasing modes appear at different wavelengths. The origin of such divergence lies in the fact that the range of collection angles overlaps only slightly, i.e. the ellipse collects angles in the range $\Theta_{1,wg} < \Theta_1 < \Theta_{1,ir}$ and the microscope only in the range between 0 and the collection angle of the microscope O2 ($\alpha_{obj}$=24°, dashed blue curve of Figure 3). Consequently, lasing modes that are efficiently collected at FPS are weak or even absent at FPW and vice versa. The inset of Figure 4a illustrates the previous statement. The logarithmic vertical scale of the graph evinces that a lasing mode that is collected efficiently at FPS is much less intense than others when collecting at FPW.

On Figure 4c we plot the intensity of the most intense lasing mode as a function of the photon flux for microspheres deposited over an ellipse (black curve) and over a waveguide (red curve). Remarkably, the ellipse configuration presents a lower lasing threshold and much higher emitted intensity. This behavior is not associated to the specific characteristics of the microspheres, which behave similarly to the black curve when measured in an isolated configuration (deposited over a $SiO_2$ substrate) [9]. Thus, we can conclude that microspheres placed on the holes created over waveguides suffer higher scattering losses than those placed over ellipses. We think that the protuberance on the hole floor associated to the presence of the buried waveguide is the responsible of such an effect. Indeed, the microsphere is probably misaligned with respect to the center of the waveguide and stands over two points (one at the hole top boundary and the other at the highest part of the hole ground, see insets of Figure 4c). Therefore, we believe that a careful optimization of a mechanical polishing process is mandatory if the waveguide configuration is to be exploited. However, as it is described in detail in Ref. [17], it is not straightforward to obtain optimum results from this process. This issue is, evidently, not present over the ellipse, hence no further optimization of the fabrication method is required in this regard.

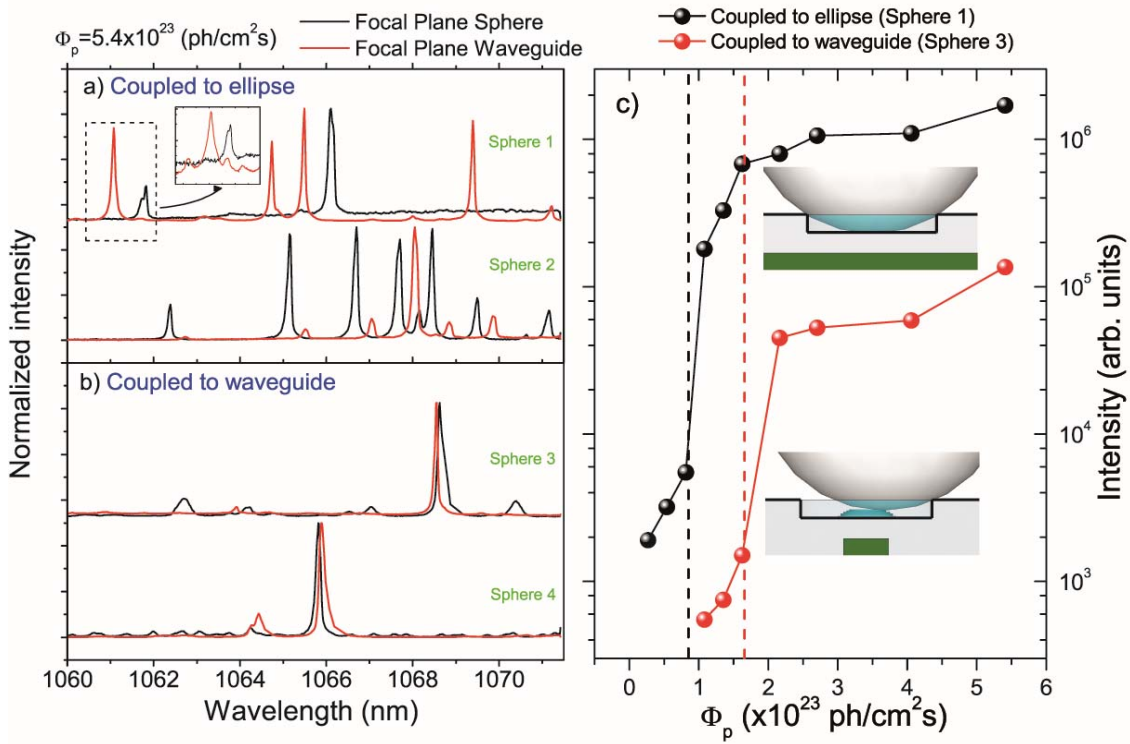

**Figure 4.** Spectra of different microspheres in a lasing regime under a pump photon flux of $5.4 \times 10^{23}$ (ph/cm$^2$s) vertically coupled to ellipses (Panel a)) and waveguides (Panel b)). The black (red) spectra are collected with the objective O2 placed at FPS (FPW). In the inset of panel a) we plot a part of the collected spectra in logarithmic scale. c) Emitted intensity of a lasing mode as a function of the photon flux for a microsphere coupled to an ellipse (black) and to a waveguide (red). A scheme of the possible accommodation of the microspheres inside the hole for each case is also included.

## VI. CONCLUSIONS

The results presented in this work demonstrate that the elliptical configuration is a reasonable strategy for coupling lasing microspheres to a photonic platform. In fact, this approach presents lower fabrication complexity than the more conventional waveguide configuration, since it allows skipping a planarization step. Above a specific value of the ellipticity (e=0.82 in the case of using $Si_3N_4$ and $SiO_2$ as the high and low refractive index materials respectively) it also shows higher light collecting capacity, achieving 50% of improvement for the structures fabricated in the current work (e=0.9) and potentially reaching 100% for ellipticities close to unity. This enhancement could be exploited in sensing applications, given that light collecting capacity is directly related with the microsphere surface able to be tested in the optical measurements. A straightforward strategy to improve further the performance of this configuration would be to increase the refractive index contrast between the ellipse and the surrounding medium. A more complex upgrading would be that of integrating a Distributed Bragg Reflector (DBR) in the back part of the ellipse, where the reflectivity of the standard ellipse is quite low.

## ACKNOWLEDGEMENTS

This work was supported by the EC through the Advanced Grant SOULMAN (ERC-FP7-321122) and the Spanish Ministry of Economy and Competitiveness through the project LEOMIS (TEC2012-38540-C02-01). JMR acknowledges the financial support of Secretariat for Universities and Research of Generalitat de Catalunya through the program FI-DGR 2013. We acknowledge A. Pitanti for fruitful discussions and for a critical reading of the manuscript.

**REFERENCES**

1. T. J. Kippenberg, J. Kalkman, A. Polman, and K. J. Vahala, Phys. Rev. A **74**, 051802(R), (2006).
2. S. L. Mccall, A. F. J. Levi, R. E. Slusher, S. J. Pearton, and R. A. Logan, Appl. Phys. Lett. **60**, 289–291 (1992).
3. A. W. Fang, R. Jones, H. Park, O. Cohen, O. Raday, M. J. Paniccia, and J. E. Bowers, Opt. Express **15**, 2315 (2007).
4. A. Polman, B. Min, J. Kalkman, T. J. Kippenberg, and K. J. Vahala, Appl. Phys. Lett. **84**, 1037–1039 (2004).
5. M. L. Gorodetsky, and V. S. Ilchenko, J. Opt. Soc. Am. B, **16**, 147-154, (1999).
6. H. Fan, S. Hua, X. Jiang, and M Xiao, Laser Phys. Lett **10**, 105809, (2013).
7. M. Cai, O. Painter, K.J. Vahala, P.C. Sercel, Opt. Lett., **25**, 1430-1432, (2000).
8. L. L. Martín, D. Navarro-Urrios, F. Ferrarese Lupi, C. Perez-Rodríguez, I. R. Martín, J. Montserrat, C. Dominguez, B. Garrido, and N. Capuj, Laser Phys., **23**, 75801 (2013)
9. J. M. Ramirez, D. Navarro-Urrios, N. E. Capuj, Y. Berencen, A. Pitanti, B. Garrido, and A. Tredicucci, arXiv:1504.03116 (2015).
10. L. He, Ş. K. Özdemir, and L. Yang, Laser Photon. Rev. **7**, 60-82 (2013).
11. L. Ding, C. Belacel, S. Ducci, G. Leo, and I. Favero, Appl. Opt. **49**(13), 2441 (2010).
12. G. S. Murugan, M. Zervas, Y. Panitchob, and J. S. Wilkinson, Opt. Lett. **36**, 73-75 (2011).
13. G. R. Elliott, D. W. Hewak, G. S. Murugan, and J. S. Wilkinson, Opt. Express **15**, 17542–17553 (2007).
14. F. Ferrarese Lupi, D. Navarro-Urrios, J. Rubio-Garcia, J. Monserrat, C. Dominguez, P. Pellegrino, and B. Garrido, J. Lightwave Technol., **30**, 169 (2012).
15. D. Navarro-Urrios, M. Baselga, F. Ferrarese Lupi, L. L. Martín, C. Pérez-Rodríguez, V. Lavin, I. R. Martín, B. Garrido, and N. E. Capuj, J. Opt. Soc. Am. B **29**, 3293-3298, (2012).
16. J. E. Geusic, H. M. Marcos, L. G. Van Uitert, Appl. Phys. Lett. **4**, 182-184 (1964).
17. M. Ghulinyan, R. Guider, G. Pucker, and L. Pavesi, IEEE Photon. Technol. Lett. **23**, 1166 (2011).